\RequirePackage[prologue,table]{xcolor}
\documentclass[acmlarge,timestamp]{acmart}

\setcopyright{rightsretained}
\copyrightyear{2021}
\acmYear{2021}
\acmDOI{10.XXXX/XXXXXXX.XXXXXXX}

\acmConference[PLoP'21]{PLoP'21: Pattern Languages of Programs 2021}{October 05--07, 2021}{HILLSIDE}
\acmBooktitle{PLoP'21, OCTOBER 5--7, HILLSIDE. Copyright 2021 is held by the author(s)}
\acmPrice{15.00}
\acmISBN{978-1-4503-XXXX-X/18/06}

\usepackage{graphicx}
\usepackage{grffile}
\usepackage{longtable}
\usepackage{wrapfig}
\usepackage{rotating}
\usepackage[normalem]{ulem}
\usepackage{textcomp}
\usepackage{capt-of}
\usepackage{hyperref}
\usepackage{mdframed}
\usepackage{afterpage}

\usepackage{booktabs}

\usepackage[pagewise]{lineno}

\usepackage{natbib}
\usepackage{float}
\usepackage{xypic}
\usepackage{tikz}

\usepackage{placeins}
\newenvironment{echo}{}{}
\usepackage{enotez}
\renewcommand{\endnote}[1]{}
\newcommand{\ignorelink}[2][]{#2}
\newcommand{\markbf}[1]{\textsuperscript{\textbf{#1}}}
\setenotez{counter-format = alph, mark-cs = \markbf}
\DeclareInstance{enotez-list}{sverre}{paragraph}{heading={},notes-sep=\baselineskip,format=\normalsize\normalfont\raggedright\leftskip1.8em,number=\makebox[0pt][r]{#1.\ }\ignorespaces,}
\usepackage{epigraph}
\date{}
\title{Patterns of Patterns: A Methodological Reflection on the Future of Design Pattern Methods}
\hypersetup{
 pdfauthor={Joseph Corneli et al.},
 pdftitle={Patterns of Patterns},
 pdfkeywords={},
 pdfsubject={},
 pdfcreator={Emacs 28.0.50 (Org mode 9.5)}, 
 pdflang={English}}

\begin{document}

\title{Patterns of Patterns: A Methodological Reflection on the Future of Design Pattern Methods}

\author{Joseph Corneli}
\authornote{Corresponding author, jcorneli@brookes.ac.uk.}
\email{jcorneli@brookes.ac.uk}
\orcid{1234-5678-9012}
\author{Alex Murphy}
\email{19032710@brookes.ac.uk}
\affiliation{%
  \institution{Oxford Brookes University}
  \streetaddress{Gipsy Lane}
  \city{Oxford}
  \country{UK}
  \postcode{OX3 0BP}
}

\author{Raymond S. Puzio}
\email{rsp@hyperreal.enterprises}
\author{Leo Vivier}
\email{zaeph@zaeph.net}
\affiliation{%
  \institution{Hyperreal Enterprises Ltd}
  \streetaddress{81 St Clement’s St}
  \city{Oxford}
  \country{UK}
  \postcode{OX4 1AW}}

\author{Noorah Alhasan}
\affiliation{%
  \institution{LBJ School of Public Affairs, The University of Texas at Austin}
  \streetaddress{P.O. Box Y}
  \city{Austin}
  \state{TX}
  \country{USA}
  \postcode{78713-8925}}
\email{noorah.alhasan@utexas.edu}

\author{Vitor Bruno}
\affiliation{%
  \institution{Milestone English}
  \streetaddress{Rua Trieste 170, ap2}
  \city{Palhoca}
  \state{SC}
  \country{Brazil}
  \postcode{88132-227}}
\email{chief@milestoneenglishcourse.com}


\author{Charlotte Pierce}
\affiliation{%
  \institution{Pierce Press}
  \streetaddress{PO Box 206}
  \city{Arlington MA}
  \country{USA}
  \postcode{02476}
}
\email{charlotte.pierce@gmail.com}

\author{Charles J. Danoff}
\affiliation{%
  \institution{Mr Danoff’s Teaching Laboratory}
 \streetaddress{PO Box 802738}
 \city{Chicago}
 \state{IL}
  \country{USA}
  \postcode{60680}}
\email{contact@mr.danoff.org}

\renewcommand{\shortauthors}{Corneli et al.}

\begin{abstract}
This paper shows how we combine and adapt
methods from elite training, future studies, and
collaborative design, and apply them to address significant problems
in social networks.  We focus on three such methods:
we use Project Action Reviews to implement social perception, Causal Layered
Analysis to implement social cognition, and Design Pattern Languages
to implement social action.  We present the results of two studies:
firstly, we use Causal Layered Analysis to explore the ways in which the design
pattern discourse has been evolving.
Secondly, to illustrate the three methods in combination,
we develop a case study, showing how we applied the methods to
bootstrap a distributed cross-disciplinary research seminar.
Building on these analyses, we
elaborate several scenarios for the future use of design patterns in
large-scale distributed collaboration.
Our case study suggests ways in which progress could be made towards realizing these scenarios.
We conclude that the
combination of methods is robust to uncertainty, insofar as they support adaptations as circumstances change, and incorporate diverse perspectives.  In
particular, we show how methods drawn from other domains enrich and
are enriched by design patterns; we believe the analysis will be of
interest to all of the communities whose methods we draw upon.
\end{abstract}




\maketitle


\section{Introduction}
\label{sec:org195e8e3}
\label{Introduction}

Our paper presents a novel synthesis of existing methods from the
fields of Learning Management, Future Studies, and Design Patterns.
We show how these methods complement each other, and combine
holistically into a coherent framework for collaborative design.

The contribution of the paper is simultaneously theoretical and
practical.  As a stepping stone to the fully integrated method
introduced here, we draw on an approach from Future Studies to analyze
existing Design Pattern literature and practices, and develop several
scenarios that characterize potential directions for the future
development of design pattern methods.  Additionally, we present a
practical case study in the new combined methodology.  A limitation of
the work presented here is that, so far, we have only anecdotal
evidence for the method’s efficacy.  However, we explain why we think
further experiments with this approach will be useful both to design
pattern thinkers, and other communities.  Indeed, we contend that an
approach along the lines developed here is needed for tackling the
world’s largest problems.



\section{Background}
\label{sec:org7c32ecc}

In 1999, the architect Christopher Alexander discussed his work with
an audience of programmers, sharing his vision of a synthesis of
architecture and computer science that could build towards the
generation of a “living world” \cite{alexander1999a}.  Considering the
existential risks we face, we may take these remarks as more than just
a metaphor.

\emph{Design patterns} offer a point of entry into this vision.  We
begin by recalling that  Alexander thought about patterns
in both a fundamental and a methodological sense.

\begin{quote}
\emph{As an element in the world}, each pattern is a relationship between a
certain context, a certain system of forces which occurs repeatedly in
that context, and a certain spatial configuration which allows these
forces to resolve themselves.\medskip

\emph{As an element of language}, a pattern is an instruction, which shows
how this spatial configuration can be used, over and over again, to
resolve the given system of forces, wherever the context makes it
relevant. \citep[p.~247]{alexander1979a} (our emphasis)
\end{quote}

Patterns in the first sense are basically physical in nature.
\emph{Design} comes into play with the second sense.  Leitner
summarized how this is meant to work: ``Patterns are shared as
complete methodic descriptions intended for practical use by experts
and non-experts'' \cite{leitner2015a}.  Already there are a number of
practical texts that use patterns (in the second sense) to talk about
patterns (again in the second sense): they share methods that aid in
discovery, writing, workshopping, and the broader application of
design patterns.  By contrast we develop a more fundamental analysis,
and use this to work towards a new level of practicality.


Alexander’s hopeful stance currently comes up against complex global
crises.  How are we to understand design patterns in this context?
Stephen Batchelor writes: “If I am to take this crisis with the
seriousness I feel it deserves, then I need to align my thoughts and
actions. I require a coherent worldview to provide a rational and
ethical foundation for my behavior” \cite{batchelor2020embracing}.
Pattern methods could fill part of this need.  It would be overly
simplistic to see in patterns only evidence of a “technical mindset”
(\emph{ibid.}), embodied only in a growing repository of technical
fixes.  Understood in their fundamental sense, patterns can help to
articulate “forms of collective action that can respond to the climate
emergency that threatens life on Earth” (\emph{ibid}.).  Now the
relevant forces are no longer simply physical, but are socially
distributed and culturally determined.


However, outstanding criticisms show that design pattern methods have
not yet reached their envisioned potential \cite{dawes2017a}.  As a
step towards that realisation, we now consider two additional methods,
and revisit the core features of the design pattern method.  The
specific selection of methods is grounded in what we have found
practical in our shared projects, though for context we provide brief
pointers to other methods that serve similar purposes.

\subsection{Project Action Review}
\label{sec:orgc354a47}
\label{par_method}

The US Army developed a methodology called the \emph{After Action Review} [AAR], which they use in training elite soldiers \cite{Training-the-Force}. Conducting an AAR facilitates learning from past experience, to generate better future performance. The method has also been used effectively in business settings \cite{learning-in-the-thick-of-it}. AARs can be used to assign responsibility when things go wrong, and can help people figure out how to do better next time. As such, the AAR shares common ground with the \textsc{Daily Scrum} and \textsc{Sprint Retrospective} \cite{sutherland2019a} patterns from the Scrum methodology.  However, it does not have the product orientation of Scrum.

In a distributed peer-to-peer collaboration, we wanted an adaptation of the
AAR which would make it more open ended and horizontal in nature.  We
came up with the following template:

\begin{enumerate}
\item Review the intention: what do we expect to learn or make together?
\item Establish what is happening: what and how are we learning?
\item What are some different perspectives on what’s happening?
\item What did we learn or change?
\item What else should we change going forward?
\end{enumerate}

When we fill in the template, we call it “doing a \emph{PAR.”}  As an acronym, “PAR” has stood for various things over the years—Paragogical Action Review,\footnote{\url{http://ceur-ws.org/Vol-739/paper\_5.pdf} (p. 5) and \emph{Peeragogy Handbook} v3, p. 134} Peeragogical Action Review,\footnote{\url{https://github.com/Peeragogy/Peeragogy.github.io/wiki/Monthly-Wrap:-March-2020}} Project Action Review—but we like PAR as a stand-alone term. Allusively, it brings to mind the corresponding concept of \emph{par} in golf, and helps give us a sense of how we are doing at any given point in time.\footnote{“In golf, \emph{par} is the predetermined number of strokes that a proficient golfer should require to complete a hole, a round (the sum of the pars of the played holes), or a tournament (the sum of the pars of each round).”—Wikipedia}\textsuperscript{,}\footnote{\url{https://web.archive.org/web/20150909224638/http://metameso.org/\~joe/docs/The-Paragogical-Action-Review.pdf}} Like the Army, we typically use PARs retrospectively (“what \emph{did} we expect to learn or make together?”).  However, PARs can also be applied to look forward (to prePARe, so to say) as a way to scaffold anticipation by “remembering the future” \cite{arnkil2008remembering}. In that case, item (5) can be expanded to include a number of alternative scenarios.

\subsection{Causal Layered Analysis}
\label{sec:org088a930}
\label{CLA_patterns}

Sohail Inayatullah developed Causal Layered Analysis (CLA) \cite{inayatullah1998b,inayatullah2004causal} as a research methodology for examining a topic of concern at four layers that he refers to as
the \emph{litany}, \emph{system}, \emph{worldview} and \emph{myth}.  Part of the reason to carry out such
an analysis is that there are different kinds of causes, ranging from
immediate events to deep-seated cultural beliefs.  Inyatullah’s work draws on his
scholarship of P. R. Sarkar:
\begin{quote}
For Sarkar, there have been four historical ways humans have dealt
with their physical and social environment: either by being dominated
by it, by dominating it through the body, dominating it through the
mind, or dominating it through the environment
itself. \cite{inayatullah1999situating}
\end{quote}
In developing a CLA, none of the four layers is
privileged over the others, nor are they examined in isolation.
Rather, one moves between them, examining how they relate to one
another.  One can then integrate these insights to form a more
comprehensive basis for understanding what is happening in the present
and for anticipating the future.  Table \ref{tab:cla-summary} describes each of the four layers according to the following schema:

\begin{itemize}
\item \textbf{Contents}: \emph{What is found in this layer?}
\item \textbf{Analysis}: \emph{Techniques for analysis of this layer.}
\item \textbf{Literature}: \emph{Instances of texts which are typically operative at this layer.}
\end{itemize}

Clemens \cite{clemens2020asian} relates CLA to the work of Gregory
Bateson: specifically, he finds an analogue in Bateson’s notion of a
“pattern of patterns” \cite{bateson1979mind} which interrelates
phenomena across disparate domains.


\begin{table}[t]
\begin{tabular}{c}
\textbf{Litany}\\
\begin{minipage}{\textwidth}
\begin{description}
\item[Contents:] Observable facts, events, and quantitative trends.
\item[Analysis:] Minimal processing of data.
\item[Literature:] News reports, tax filings, chit-chat.
\end{description}
\medskip
\end{minipage}\\
\textbf{System}\\
\begin{minipage}{\textwidth}
\begin{description}
\item[Contents:] The social, economic, political, and historical forces which shape events.
\item[Analysis:] Technical explanations and interpretation of data within a given paradigm.
\item[Literature:] Editorials and policy institute reports.
\end{description}
\medskip
\end{minipage}\\
\textbf{Worldview}\\
\begin{minipage}{\textwidth}
\begin{description}
\item[Contents:] Core values and attitudes which motivate choices and actions.
\item[Analysis:] Uncover deep assumptions and study the mental and linguistic constructs which undergird how people interact with each other and their surroundings.  Compare and critique paradigms and discourses.
\item[Literature:] Works of philosophy and critical theory.
\end{description}
\medskip
\end{minipage}\\
\textbf{Myth}\\
\begin{minipage}{\textwidth}
\begin{description}
\item[Contents:] The symbols and tales which give meaning to life.
\item[Analysis:] Study symbols and narratives, and the myths and rituals within which they participate.
\item[Literature:] Poetry, art, anthropology, Jungian analysis.
\end{description}
\end{minipage}
\end{tabular}
\medskip
\caption{Overview of the layers in Causal Layered Analysis\label{tab:cla-summary}}
\vspace{-2\baselineskip}
\end{table}

\subsection{Design Pattern Languages}
\label{sec:org5987433}
\label{dpl_method}

The two senses of ‘pattern’ mentioned above—‘As an element in the \emph{world}\ldots{}’ and
‘As an element of \emph{language}\ldots{}’—are mirrored within the
concept of a design pattern.  Like an ellipse, the design pattern has
two main foci: context and community.

\begin{itemize}
\item \emph{Context} shapes and constrains the type of activity which is being considered, be it designing a building, writing software, or something else.
\item \emph{Community} encompasses the stakeholders—experts and non-experts alike—who are involved with or otherwise affected by a particular project.
\end{itemize}

Integral to the basic concept of a design pattern is a third feature
that describes the interaction of the community and the context.  The
community uses the pattern to overcome some real or potential \emph{conflict}
that they experience within this context.  It bears emphasis that the
community is not assumed to be homogeneous, and, indeed, this may be part of how
the conflict is experienced; i.e., it need not be the case
that all members of the community share the same experience or view of
the context, nor that they are all uniformly affected by
the circumstances arising therein.  The conflict is also referred to as a \emph{problem}; its resolution is
described as a \emph{solution}.  Alexander and Poyner emphasized that
‘design’ is not needed when the conflict can be resolved in an obvious
or straightforward manner.  For example, you typically would not need
a design process surrounding \emph{sitting in a chair},
\begin{echo}
because “under normal conditions each one of the
tendencies which arises in this situation can take care of itself”
\citep[p.~311]{alexander1970a}.\endnote{The straightforwardness of sitting in a chair notwithstanding, Thich Nhat Hahn has written a book called \emph{\href{https://www.penguin.co.uk/books/111/1111997/how-to-sit/9781846045141.html}{How To Sit}} (2014): this somewhat proves Alexander’s point as the exception to the rule.  However, prior to reading this book one might want to read /How to Read a Book/.}
\end{echo}

We might say that the design pattern carries with it a fragment of
irreducible complexity.  This perspective may or may not be
surprising. Early on, Alexander described the need for patterns when things get complex
\cite{alexander1964notes}.  He specifically focuses on what could be
called “horizontal” complexity, a situation where there are a lot of
moving parts and relations between them.  Methodologically this is
elaborated with the notion of a \emph{pattern language}.\endnote{The issues involved become somewhat more complex when there are multiple DPLs interoperating, but are not fundamentally different.}
Pattern languages have a property of unfolding, from more general to
more specific.  However, they do not necessarily cover deeper forms of
“vertical” complexity, where there are deep historical or ontogenetic
causes, feedback loops, or complex conceptual issues which are not readily
expressible in design-pattern-theoretic terms.  Let’s have another
look at these issues by way of two contrasting metaphors.

The first metaphor comes from Christian Kohls, who proposed to treat
each design pattern as a journey: “a path as a solution to reach a
goal” \cite{kohls2010a}.  In this metaphor, design patterns are
understood to have an initial condition and an end condition, defined
within some context. The context also associates a cost to traversals
of paths.  There are several associated problems: the elementary
problem is to traverse the terrain and travel from the start state to
the end state.  The next problem is to do this at low cost.  The third
problem is to find a reliably repeatable way to do this.  A fourth
problem is to describe the process in such a way that the path can be
traversed by others.

The second metaphor comes from Joseph Campbell, who described an
“archetypal pattern” \cite{shalloway2005a}, one that can be found
embedded in myths and stories across diverse cultures and historical
periods.  The “hero’s journey” is also described with a path
\cite{campbell1949a}, however, in this case the path runs in a circle,
and the journey focuses on the transformations of the hero who
traverses it.  Although an account of the journey can be shared,
traversal is effectively single-use.  The cost is typically “high.”
Nevertheless, once a myth or metaphor is established in a shared
narrative, the journey can be reenacted through ritual or engaged with
in other ways that solve a range of social problems
\cite{handelman1998a}. In short, the difference between these two
traversal stories suggests that the process of finding “the path that
is capable of leading to a good structure” \cite{alexander1999a} may
contain irreducible complexity—even when sharing the information
about the path is relatively simple.

\section{Methodology}
\label{sec:org134acbb}
\label{methods}

Each of the three methods described above has distinct use-cases when
considered in isolation.  Viewed as alternatives, they would have
various tradeoffs between them, concerning their ease of use, their
ability to generate solutions, the breadth of their applicability, and
so on.  In this section we put them together as one holistic pattern
of patterns.  We will use them to scaffold social perception,
cognition, and action (Table \ref{tab:acronyms}).

\begin{echo}
\begin{table}[h]
\begin{tabular}{llll}
\emph{Key verbs:}           & perceive       & think            & act\\[.2cm]
\emph{Scientists refer to:}& “sensory” & “cognitive” & “motor” systems\\[.2cm]
\emph{Our implementation:}&
\begin{minipage}{1in}
\textbf{P}roject\newline
\textbf{A}ction\newline
\textbf{R}eview
\end{minipage}&
\begin{minipage}{1in}
\textbf{C}ausal\newline
\textbf{L}ayered\newline
\textbf{A}nalysis
\end{minipage}
&
\begin{minipage}{1in}
\textbf{D}esign\newline
\textbf{P}attern\newline
\textbf{L}anguages
\end{minipage}
\end{tabular}
\vspace{.5cm}
\caption{Three acronyms used in this paper: PAR, CLA, and DPL\label{tab:acronyms}}
\end{table}
\end{echo}

We adopt this tripartite division from classical psychology
\cite{Hilgard1980}; it continues to be relevant in contemporary
neuroscience \cite{Teufel2020,Friston2013}.  Indeed, it is not only as
a division, but precisely in characteristic combination that these
three factors become important.  This suggests a third understanding
of ‘pattern’ as “a dynamic [...] ongoing awareness created and upheld
by a mass of ongoing physiological activity” \cite{iran2000bartlett}.
To study such factors in integration “requires understanding the
conditions and laws of construction in mental life” (\emph{ibid.}).
So it is, as well, for collective knowledge production.  To
recapitulate and reframe the methods in these terms:

\paragraph{Project Action Review (PAR) is the sensory element: systematically gathering and verifying observational data.}
\label{sec:orga7a1581}
The Project Action Review is structured around five questions which
members of a community discuss and answer together.  This practice
generates a record of an event, as seen through the eyes of the
participants.  In the moment, the PAR allows us to draw out views
which might have gone unstated otherwise.  Over time, projects which
use PARs improve their chances of staying grounded in reality as
circumstances evolve.  We can additionally use PARs to help check how
effectively we are using other methods.

\paragraph{Causal Layered Analysis (CLA) is the cognitive element: giving organization and depth to the enterprise.}
\label{sec:orgf7e1393}
The goal of this methodology is to achieve a deep and inclusive
understanding by integrating empiricist, interpretative, critical, and
actionable knowledge surrounding a topic of concern.  Without an
integrated understanding, a group runs the risk of getting lost in a
muddle of details.  CLA can pull information logged in PARs
together into a coherent body of self-knowledge.  It can also help to surface
concerns that might remain implicit in pattern language: for example,
CLA could help us understand why we had prized a technological
solution to what, upon consideration, turned out to be a fundamentally
social issue (or vice-versa).

\paragraph{Design Pattern Language (DPL) is the motor element: orchestrating and scaffolding action}
\label{sec:org4462a85}
Having carefully analyzed the situation and identified possible
solution pathways, we must plot a course of action that accounts for the
complexities of the situation.  Individual design patterns present
solutions to recurring problems: they can be combined with other
patterns and adapted to different situations.  A DPL is a common
language which a community can use to discuss matters of design; it
serves as a repository of shared knowledge.  The flexibility of DPLs
allows the structure to be customized to our particular circumstances
as they evolve.

\subsection*{PLACARD: A Synthesis of PAR, CLA, and DPL}
\label{sec:orgd924cd6}
\label{methods_summary} We are now in a position to explain how PAR, CLA,
and DPL combine into one holistic pattern, in Leitner’s sense of a
complete methodic description \cite{leitner2015a}.  We will write this
down using the classic DPL format: describing the associated
\emph{context}, the \emph{problem} denoting a conflict, together with a \emph{solution}.
As it happens, the three acronyms introduced earlier can be combined and remixed
to provide a title for this pattern.
$$\textrm{PAR}+\textrm{CLA}+\textrm{DPL}=\textrm{PLACARD}$$
This accurately suggests that
the methods need not be run in a fixed order, but are interwoven together.
\label{sec:org958ff04}
\label{PLACARD}

\begin{itemize}
\item \textbf{Context}: In the course of working on a project: \emph{we use the PAR to get a sense of our working context}.
\item \textbf{Problem}: Although we may encounter many difficulties in this context, our effort to understand them faces a central \textbf{challenge}, namely the fact that the problems span different layers and scales of complexity, so it can be hard to understand where the difficulties actually come from: accordingly, \emph{we use the CLA to understand and frame the problems and their interconnections}.
\item \textbf{Solution}: Once we have grasped the problem, we need to elaborate an actionable solution that remains adaptable to ongoing changes in the context: \emph{we use DPL to elaborate the solution (returning to PAR and CLA as needed)}.
\end{itemize}

\begin{figure}
\centering
\includegraphics[width=.5\textwidth]{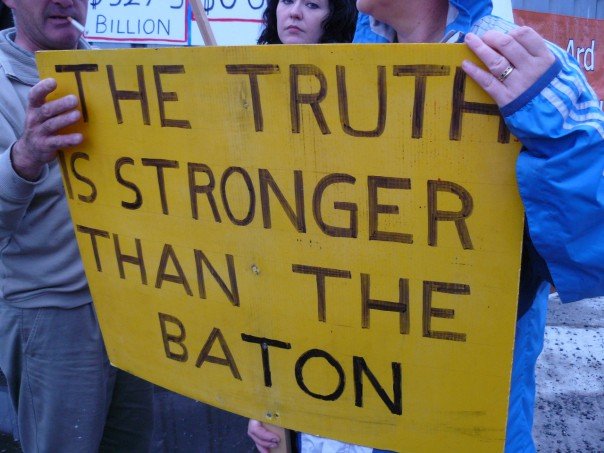}
\caption{\label{fig:placard}Mnemonic illustration of the \ignorelink[PLACARD]{PLACARD} pattern
}
\Description[People holding a placard]{People holding a placard that reads “THE TRUTH IS STRONGER THAN THE BATON”}
\end{figure}

Figure \ref{fig:placard} provides a mnemonic.\footnote{For French speakers, \emph{placard} means ‘cupboard’, and there is an idiom, \emph{placardisé}, which refers to an employee whose tasks all have been reassigned to others; the import is similar to the English idiom ‘put to pasture’. While it is not the case that \ignorelink[PLACARD]{PLACARD} reassigns all DPL functions to other methods, the French idiom is potentially suggestive as another mnemonic.} The main thing to notice is that using the three methods
together can help make the design pattern method practicable.  We can use the
PAR to move from a context to a “Context”, established and written
down.  We can use CLA to move from a situation of concern to a
situation in which the core “Problem” or “Problems” can be thought
about.

The fact that DPL shows up inside of \ignorelink[PLACARD]{PLACARD} may be somewhat concerning.
\begin{echo}
The reader may be wondering: “I think I can see how the methods that have been discussed could help in understanding
the \textbf{context} and the \textbf{problem}, but is there anything here that can actually help formulate \textbf{solutions}?”\endnote{\textbf{JC@coauthors:} I’m still a bit concerned about this!}
\end{echo}
This concern calls for some further brief remarks before turning to a
case study of the \ignorelink[PLACARD]{PLACARD} pattern in use.

A solution process can often be decomposed into interrelated subtasks
\cite{alexander1964notes,polya1945solve,polya1954plausible}.  A
standard problem-solving methodology is to understand the
\mbox{(sub-)}contexts and \mbox{(sub-)}problems in detail—along with
their relationships to other parts of the developing decomposition—and
on this basis make predictions about the way an intervention could
change the overall system.  Accordingly, problem solving often goes
back and forth between “two different forms of information processing:
bottom-up and top-down”\footnote{(Punctuation modified.)} \cite{Teufel2020}.  CLA
can help with thinking in both directions.  Nominally, CLA is an \emph{analytic} tool that
decomposes a problematic situation into \emph{layers}, and
\emph{causes} operating at and between these layers: in this sense it
functions top-down.  However, CLA also plays a bottom-up synthesis role.
Whereas we could compare the PAR to a tangent vector or derivative
that gives a momentary reading of how things are going at a given
point in time, CLA can be used to integrate these observations into a
plan.

\section{Results}
\label{sec:orgac68a6d}
\label{Results}




We report on two studies traversing different scales.  Firstly, we
applied CLA to the design pattern literature and practices, with the
purpose of scaffolding an examination of the future of the design
pattern theory.  Secondly, we made practical use of CLA alongside the
PAR and DPL methods (i.e., PLACARD as a whole) within a case study of
a distributed research seminar.
\begin{echo}
Details of these
analyses are presented in Appendix \ref{Analysis} and Appendix
\ref{Case_study}, respectively, henceforth referred to here as the
\textbf{\hyperref[Analysis]{Analysis}} and \textbf{\hyperref[Case_study]{Case Study}}.
\end{echo}

\subsection{Analysis: Design Pattern Language literature and practices}
\label{sec:orgca77803}

In this study we examined DPL literature and practices in the form of books,
articles and conferences.  In order to counterbalance the coverage, we
included dissenting and critical voices.  Here we did not have PARs at
our disposal, so the sensory element was provided by the views
expressed by previous authors.

In précis, we were interested in the following topics:
\begin{itemize}
\item Queries raised by Alexander and his collaborator Bryant, along with
a systematic analysis of criticisms of pattern methods collected by
Dawes and Ostwald.
\item Issues related to how people share and discuss patterns, as well as
the changing way in which these discussions have been framed at
PLoP.
\item The worldview linked with patterns through the lens of mob software
and its critiques.
\item Symbols and philosophical traditions that enrich our understanding
of the context in which Alexander developed his methods.
\end{itemize}

Taking a deep dive into DPL via CLA allowed us to gain perspectives on
how design patterns work.  In particular, we have illustrated the
complexity that underpins the model.  Alexander expands on his
metaphysical considerations in his multi-volume work, \emph{The Nature of
Order} (\emph{TNO}).  Our Analysis shows how some of these more
ephemeral-seeming factors are ramified across various layers of the
pattern theory.

In particular, working from the myth layer towards the more surface
layers: we show that the notion of \emph{wholeness} that Alexander deploys is
complex, and that the corresponding theory of emergent order based on
this concept is similarly complex and linked to “a tension between
independent and conforming tendencies” \cite{vandrunenchristian}.  This
tension has—only partially—resolved into a dichotomous relationship
between \emph{users} and \emph{designers} of patterns and pattern-linked artifacts.
There are however many remaining points of friction, as users of
pattern methods run into difficulties \cite{dawes2017a}, and designers
are not entirely clear on how to improve the situation.

\subsection{Case Study: Emacs Research Group}
\label{sec:org70a9ab2}

The Emacs Research Group (ERG) is a transdisciplinary seminar organized
around the theme of research in, on, with, and about the Emacs computer program.\footnote{\url{https://www.gnu.org/software/emacs/}}
Although Emacs is best known as a text editor, its extensibility and
self-documenting nature make it a more general platform for dealing with
symbolic content, and a vibrant site for research into writing and programming.
ERG aims to explore topics such as the following:
\begin{itemize}
\item How is Emacs \emph{used} to help conduct research in various disciplines?
\item What is the \emph{user experience} with Emacs and what is the user community?
\item How could the Emacs system \emph{interoperate} with other computer programs?
\item How might the \emph{communities} affiliated with Emacs interact with other communities of software users and developers?
\item What new \emph{functionalities} would broaden the applicability of Emacs?
\end{itemize}
ERG meets approximately weekly, sometimes inviting guest speakers.  After each
meeting, the participants summarize their experience in a PAR.  This
serves to surface matters of concern and highlights in each session of
the seminar.  Every six meetings, the techniques of CLA were used to
organize and condense the PARs into a coherent statement of purpose.
Finally, Peeragogy design patterns are used to formulate a plan of action
informed by this analysis.  Thus, all three methods are combined in
line with the \ignorelink[PLACARD]{PLACARD} pattern introduced above.

In this study, we were most interested
in understanding and exploring the PLACARD method and its potential
efficacy; e.g., could we use the PLACARD-associated methods to develop actionable and
deliverable designs for new software?  After a year of working with
the methods together, we presented them at EmacsConf 2021, advocating
that other Emacs enthusiasts adapt them for use in creating their own
small research
groups.\footnote{\url{https://emacsconf.org/2021/talks/erg/}}

\subsection*{Summary of Findings}
\label{sec:org990bf98}
We used Causal Layered Analysis to describe the evolution of Design
Pattern Language methods in response to criticism, innovation,
technical developments, and long-term cultural change.  We
additionally surface some of the forces and tensions in the discourse.
We used Design Pattern Language methods, together with Project Action
Reviews and Causal Layered Analysis, to organize a research seminar.
This case study shows how the methods can be fruitfully combined, and
suggests how others might take up the combined methods (e.g., by
substituting ‘Patterns’ for ‘Emacs’ as the focus in a future research
seminar).

\section{Discussion}
\label{sec:org721ecf9}
\label{Discussion}

Informed by the two studies described above, we would like to reflect
on why putting the CLA and DPL methods together can make a big
difference in practical terms.  To do this, we begin by examining a
specific problem domain to which CLA and DPL have been applied
separately.

Anthropogenic climate change is a situation of major global concern in
the early 21st Century.  It comes as no surprise that it has been
examined separately by proponents of both CLA and DPL.  We use this
recent history to frame future work building on the case study and
analysis developed above.

In an overview of their work on the Cooling the Commons pattern language, Cameron Tonkinwise and Abby Mellick Lopes write:
\begin{quote}
A design pattern is first an observation: “People in that kind of designed situation tend to do this sort of thing”. It is then possible to design an intervention that redirects those tendencies. If that intervention succeeds, it can become a recommended pattern to help other designers: “If you encounter this kind of situation, try to make these kinds of interventions” \cite{theconversation2021}.
\end{quote}
They amplify the ‘ethical’ aspect of their thinking:
\begin{quote}
\ldots{} the patterns we are talking about, context-specific interactions
between people and things, are more like habits. They are tendencies
that lead to repeated actions.
\end{quote}
The 41 patterns they have developed include examples like \textsc{The Night-Time Commons},\footnote{\url{https://www.coolingthecommons.com/pattern\%20deck/}}
which:
\begin{quote}
\ldots{} might shift daytime activities into cooler night times.  Some
places already have these patterns: night markets and night-time use
of outdoor spaces.  If locally adapted versions of these patterns
encourage people to adopt new habits, other patterns will be needed.
These will include, for example, ways to remind those cooling off
outdoors in the evening that others might be trying to sleep with
their naturally ventilating windows open.  Such interlinked patterns
point to the way pattern thinking moves from the big scale to the
small.
\end{quote}
Reading this, we were concerned that, while the Cooling the Commons
patterns do acknowledge \emph{horizontal} complexity—namely, through
interlinked patterns—the process does not deal with the \emph{vertical}
complexity coming from the fact that diurnal rhythms are deeply
embedded in biology and culture.  People have cultural beliefs about
the activities that are appropriate for different times of day.
Public and domestic rituals are organized about the daily cycle.
Times of day have symbolic associations.  As far as we could tell,
these authors focused on more or less technical issues at the systems
level, and did not acknowledge these issues at the worldview and myth
levels.  A more comprehensive approach might, for instance, re-examine
rituals to see which of them relate to the phenomenon of sunrise
versus the act of getting up and starting the day, and then figuring
out how to adapt these rituals to a new schedule.  A suitable research
strategy might be to study how practices changed in the past, as with the
introduction of industrialization and its clockwork regimentation of
the day.

Meanwhile, Heinonen and coauthors \cite{HEINONEN2017101} describe a CLA game that explored four
different scenarios in small groups.  The four scenarios were “Radical
Startups”, “Value-Driven Techemoths”, “Green DIY Engineers” and “New
Consciousness”.  As groups worked through the CLA for each scenario,
they developed a range of new ideas.  We wondered, how might these CLA-linked
reflections collate against the Cooling the Commons patterns?
Might players of the CLA game have spotted ways in which the patterns would conflict
with deeper values—or ways in which they might be exploited to cause
chaos \cite{friction2016a}?

Broadening our exploration of how design patterns relate to futures
studies, we note that Schwartz \cite{schwartz1996a} (Appendix,
pp. 241-248, \emph{viz.}, his “Steps to Developing Scenarios”)
described a process that follows an outline that is strikingly similar
to a design pattern template.  Both Alexander and Schwartz advocate
the identification of driving forces in a context.  However, unlike
Alexander, Schwartz does not intend to resolve conflicts between the
forces within a harmonizing design.  On the contrary, the aim in the
scenario development method is to understand how these forces might
evolve and lead to the further diversification of scenarios.  To
simplify the process, some scenario planners reduce the number of
forces to the two most important.  For example, the scenarios of
Heinonen et al.~are organised along two axes: one spanning different
degrees of integration between peers, the other, different levels of
ecological awareness.

With these reflections in mind, we came up with four scenarios for the
future development of design patterns.  The four scenarios offered
here are inspired by the four CLA layers: litany, system, worldview,
and myth.

\subsection{PLACARD becomes transferable and computational}
\label{sec:org123b4c3}
With further work, PLACARD could become a more refined but nevertheless
easy-to-use instrument for gathering, organising, and sharing data on problem
solving in social networks.  Building on this collected data,
we could take further steps to develop computational models of future imaginaries.
Patterns have previously been discussed in explicitly computational
terms—however, that direction of work so far remains mostly at the
level of a proposal \cite{alexander1999a,moran1971a}, with only
limited discipline-specific uptake (e.g., \cite{jacobus2009a},
\cite{OXMAN1994141}, \cite{taibi2003formal}).  Could this change, to
generalize the kinds of patterns that can be computed with, and make
them interoperable?  Polya had already been writing about \emph{patterns of
plausible inference} the year that Alexander started his undergraduate
degree in Cambridge \cite{polya1954plausible}; four years later
Polya’s student Allen Newell was beginning to think about how to model
the inference process computationally
\cite{newell1958,newell1983heuristic}.  In the domain of economics,
Ostrom-style institutions are analogous to design patterns
(\cite{ostrom2009a}, p. 11). Recent work looks at how descriptions of
such institutions can be extracted from text \cite{Rice2021}.  Could
this line of thinking be extended, so that other similar kinds of
patterns could be recognized where they appear?  Could the extracted
descriptions be used directly in computations?  One fruitful strategy might be to think of design patterns as \emph{conceptual blends} \cite{Corneli2018}, which can be given a computational interpretation \cite{SCHORLEMMER2021118}.  For example, the \textsc{Community Library} pattern from the Cooling the Commons pattern language blends a learning space with a cool refuge.  Could such complexities be reasoned about computationally?

\subsection{Pattern language authoring communities move to free/libre/open source licensing}
\label{sec:org4d1dd6e}
In the field of policy, \emph{resilience} describes a society’s
ability to recover after a shock; whereas \emph{adaptive capacity} describes
its ability to move to a new state \cite{thonicke2020advancing,magnan2010better}.
This ability is, in turn, linked with the health and adaptivity of the society’s
institutions \cite{fidelman2017institutions}.  Free/Libre/Open Source licensing is
one possible institutional innovation in the way design patterns are used.
Widespread adoption of new licensing arrangements could
support social learning, which could, in turn, boost adaptive capacity \cite{THIHONGPHUONG20171}.
As an example of work heading in this direction, Mehaffy and coauthors collaborated with Ward
Cunningham to make their book \emph{A New Pattern Language for Growing
Regions} \cite{mehaffy2020new} into a wiki, \href{http://npl.wiki}{npl.wiki}, which is licensed
under CC BY-SA 4.0.  Will other pattern developers follow suit and
move to open licensing—and suitable infrastructures for working with open contents?   We
can also ask: what about other kinds and qualities of openness?
A ‘copyleft’ license is not a panacea for all ills \cite{Krowne_Puzio_2006}, and
would not on its own make the pattern theory and methods fully open in all the ways that matter.
Nevertheless, grappling with the challenges around licensing and related considerations could serve
as a rallying point for the pattern community.

\subsection{Patterns empower individuals and communities}
\label{sec:orgd9f091e}

As we’ve seen in our work with Emacs and Peeragogy (and previously
with the online community PlanetMath \cite{krowne2003,corneli-thesis}) projects need more
than just access to source code in order to thrive.  We see a link
to the topic of reproducible research.  Above and beyond the immediate
technical considerations \cite{sandve2013ten}, the process of doing
science is “reproducible” if the methods are teachable to others.  The
Literate Programming paradigm can help with this.\footnote{For notes on doing reproducible research with Emacs, see \url{https://emacsnyc.org/2014/11/03/org-mode-for-reproducible-research.html}}
  At the same time, collaboration
across different skill sets is challenging; large scale
problems like adapting to climate
change seem to require such collaboration, and almost certainly won’t be solved if we carry on doing business as usual.  In the Minnesota
2050 project, participants were selected from a variety of professions
and leadership roles to produce scenarios for energy and land use, and
combined modeling with scenario planning \cite{olabisi2010}.
Actually solving large-scale problems in interdisciplinary
teams will require new thinking and additional tools: to bridge
between the viewpoints of, e.g., professional futurists, programmers,
data scientists, local farmers—and to draw on the insights of
citizen scientists \cite{wildschut2017a}.  CLA and PLACARD
can help; additional patterns of patterns will be needed
to work fluently across domains, levels, and spheres of endeavour.

\subsection{Patterns facilitate economic empowerment}
\label{sec:org1a3e003}
Access and meaningful participation are serious matters of concern in
our current technological culture \cite{unger2019knowledge}.  Patterns
could become the basis of widespread capabilities and substantial opportunities
\cite{sep-capability-approach}, shared not only
by an elite group of hackers or a few highly-paid rockstars, but by everyone.
Patterns have been used to describe soft skills that
are useful for aspiring programming professionals
\cite{hoover2009apprenticeship}: however this falls far short of
reforming the tech sector or our broader relationship to technology.
As a related example from the world of literature,
Herman Hesse’s novel \emph{The Glass
  Bead Game} imagines a society in which the elite community of scholars
studying abstract patterns form a strong hierarchy, and
remain mostly out of touch with the practical realities experienced by outsiders.
The story serves as a warning: we must proceed with caution when we seek to bridge practice and theory.
When reflecting on futures-oriented discourses, Slaughter described a spectrum:
“participatory and open at one pole and closed (or professionalised)
at the other” \cite{SLAUGHTER1989447}.  Are pattern authors prepared
to work with widespread stakeholders to study the forces that shape the economy and society?

\section{Conclusion}
\label{sec:org47c58d9}
\label{Conclusion}

In 1999, the architect Christopher Alexander discussed the future of
design patterns with an audience of programmers \cite{alexander1999a}.  We revisited this
topic from a methodological perspective.  Building on the analysis and
case study we developed, we discussed several scenarios for the future
of design pattern methods.  Our vision for change is that these four
scenarios will be given serious thought by other members of the
patterns community.  Our Case Study suggests concrete ways in which
progress might be made towards realizing these scenarios, or, indeed,
if warranted, towards rejecting them and developing another vision.

Progress will become measurable through markers of debate and dialogue
between the different communities whose work we have drawn upon, and
potentially, through trial-and-error uptake or adaptation of the
methods we’ve described.
We are certainly not the only people to think about systems and
futures: what is distinctive about this paper is that we’ve connected
these domains with design pattern terminology and methods.
Some potential implications of that connection are embedded in
the scenarios outlined above.

In the present exploration, we began by thinking about patterns from a
fundamental perspective: \emph{patterns as elements in the world}.
Some patterns repeat in space, some in time, some in both space and
time; think of a tiling, a beat, a wave.  However, patterns cannot
repeat exactly or forever: their elements are subject to spatial or
temporal displacement, and other forms of variation.  We need suitable
abilities—and methods—to perceive and work with patterns.  The methods
we used in this paper were the Project Action Review (PAR), Causal
Layered Analysis (CLA), and Design Pattern Languages (DPL)—though
other methods that fulfill the same basic purpose could be used
without significantly changing the overall import of what we say here.
\begin{itemize}
\setlength\itemsep{1.5ex}
\item By using the PAR (or another sensory method), we are able to identify recurring themes.
\item Then, by using the CLA (or another cognitive method), we are able to organize these repeating themes in a structure that exposes the underlying trends, causes, and potential terminating states.
\item With DPL (or another motor method) we can make what we have learned actionable.
\end{itemize}
A contribution of this paper was to link the sensory/cognitive/motor theory with design patterns, which we illustrated by combining PAR, CLA, and DPL into the \ignorelink[PLACARD]{PLACARD} pattern.

A key experiential finding that emerged from our joint work on this
paper is that we need to take health, well-being, and time for
reflection into account, even more than we worry about accomplishments
or deliverables, or we are likely to run out of energy before we get
very far.  Outcome-based planning needs to be balanced
against more intention-based planning that takes human
factors into account.  In future work, we hope to explore the rich
factors that contribute to health and adaptivity in groups and networks.  The
methodological contribution of this paper can help to scaffold that
process.  Alongside the PLACARD method, we have started to develop a
set of design requirements for software that can help people apply the
method in their working contexts.

The limitations of the research presented will be clear.  Our case
study examined a small and well-integrated population, which is
moreover, a subset of the authors of the present paper.  While the
case study provides a proof of concept for the applicability of the
methods, in other settings, some or all of the methods might not be
needed, or accepted without significant alteration.
Despite the limitations of this study when
it comes to external validity,
evidence to support the use of the integrated
\ignorelink[PLACARD]{PLACARD} pattern may accumulate rapidly, if the method proves useful for other transdisciplinary collaborations.
By using and negotiating around the same DPL, disparate groups may get a useful external source of validation.
We hypothesise that PLACARD could quickly make the work of Learning Management, Future Studies, and Design Pattern practitioners more robust.

We can imagine the methods we’ve discussed being
readily applied, initially, in workshops or other small groups.  For example, if
we were to gather a group of citizens, experts, and other stakeholders
to talk together about the city, they could use the methods we’ve
described to surface issues, rehash concerns, and sensemake together.
“Anticipatory Social Science” is a broader term for this kind of work.
Moreover, working with methods that distribute perception, cognition,
and action, we may become more comfortable with uncertainty, and
better able to support innovation.  Especially when grappling with
massive problems like climate change, we will need to develop a
transformative set of tools and methods \cite{miller2018transforming};
and beyond this, a coherent worldview, and perhaps even new myths.
Citizen science has a potentially important role to play here
\cite{wildschut2017a}, alongside other forms of participatory culture.
The methods we described may help to support widespread engagement,
and coordinate activities across scales: from individuals, to communities
and social movements.  As Nietzsche said, “the more affects we are able to
put into words about a thing, the more eyes, different eyes, we can
use to observe one thing, the more complete will our ‘concept’ of this
thing, our ‘objectivity,’ be” \cite{nietzsche2017nietzsche} (p. 128).


\section*{Acknowledgements}
\label{sec:org928ca70}
We thank our PLoP 2021 shepherd Michael Mehaffy for supportive
comments and productive discussion.  We thank the participants in our
PLoP Writer’s Workshop group, and Mary Tedeschi and her students in
Information Systems Development at Baruch CUNY for additional comments
on the Conference draft.  This manuscript is distantly derived from
submissions to the Connected Learning Summit 2018, Anticipation 2019,
and the \emph{Futures} journal.  We thank Paola Ricaurte, Analua
Dutka-Chirichetti, Hermano Cintra, Lisa Snow MacDonald, and Verena
Roberts—coauthors of these earlier submissions—for helping to shape
our thinking here.  We thank Claire van Rhyn for bringing the
Anticipation conference to our attention and thereby putting us in
touch with the field of futures studies.  We appreciatively
acknowledge feedback on the ideas in the paper that we received as
they were developing, from Roland Legrand, Charles Blass, Stephan
Kreutzer, Giuliana Marques, Cris Gherhes, Qiantan Hong, Cameron Ray
Smith, David O’Toole, Rebecca Raper, Steve Corneli, and Elizabeth
Cheetham.

The image appearing in Figure 1 was made available under the terms of CC Zero by User:Lapsed Pacifist at \texttt{https://commons.wikimedia.org/wiki/File:Shell\_to\_Sea\_placard.jpg}.

\section*{Supplementary Material}
\label{sec:org4dfba23}
\begin{echo}
In order to support a concise treatment of our core findings, we supply details of our Analysis and Case Study in Appendix \ref{Analysis} and Appendix \ref{Case_study}.
\end{echo}

\appendix
\section{SUPPLEMENT: Analysis: CLA applied to Design Pattern Language literature and practices}
\label{sec:org40b3af5}
\label{Analysis}

In developing this analysis we are aided by several additional methods
from the Poststructural Futures Toolkit \cite{inayatullah1998b}.  We
will refer to the individual methods from this toolkit with their names
rendered in all-caps.

\subsection{Litany: Understanding data, headlines, empirical world (short term change)}
\label{sec:org3833701}

The first layer in CLA is the \textbf{litany layer}: it describes the problems
that people are well familiar with.  In the case of the design
patterns discourse, this level includes—in particular—the familiar
kinds of conflict-based problems that are described in patterns and discussed
at PLoP, along with higher-order problems of application, and debates about these (e.g.,
ranging from Christopher Alexander’s “\textsc{Entryway Transition}” pattern to
his remarks about how people who attempted to apply his methods
ended up placing “alcoves everywhere”, etc.). This layer is sometimes
also referred to as the \textbf{problem level}: in the patterns discourse,
problems abound.  Indeed, one of the core attributes of the pattern community is that it
is not only comfortable with problems but that it actively seeks them
out with a ‘problematizing’ discourse.

Not all of the well-known and discussed problems have been solved.
For example, ‘Alexander's
Problem’, as described by his collaborator Greg Bryant, is that:
\begin{quote}
\ldots{} despite all of the tools he created, his penetrating research, his
many well-wrought projects, and his excellent writing, he did not
manage to grant, to his readers, the core sensibility that drove the
work. He also did not organize the continuance of the research program
that revolves around this sensibility. \cite{bryant2015}
\end{quote}
Attempts to work out a practical solution to this problem are
developing.\footnote{\url{https://www.buildingbeauty.org/} and
\url{https://www.buildingbeauty.org/beautiful-software}} Coming at the same
basic issue from a more visionary standpoint, Alexander framed this
query for the programmers who were using pattern methods at the turn
of the millennium:
\begin{quote}
What is the Chartres of programming? What task is at a high enough
level to inspire people writing programs, to reach for the stars?
\cite{alexander1999a}
\end{quote}
More recently, Dawes and Ostwald \cite{dawes2017a} develop an
elegant taxonomy of criticisms of the pattern method.  In
outline, their taxonomy covers criticisms at the following three
layers:
\begin{description}
\item[{Conceptualization}] Ontology, Epistemology \newline \emph{(e.g., “Rejecting pluralistic values confuses subjective and objective phenomena”)}
\item[{Development and documentation}] Reasoning, Testing, Scholarship \newline\hfill \emph{(e.g., “The definitions of ‘patterns’ and ‘forces’ are inexplicit”)}
\item[{Implementation and outcomes}] Controlling, Flawed, Unsuccessful \newline\hfill \emph{(e.g., “Patterns disallow radical solutions”)}
\end{description}

By showing how the criticisms relate to one another, Dawes and Ostwald
begin to develop a \ignorelink[GENEALOGY]{GENEALOGY} at the level of critical perspectives.
The critiques they examine show that there is not just one pattern
discourse, but many.  In a parallel work the same authors analyze the
structure of Alexander’s classic text, \emph{A Pattern Language} (\emph{APL}) and develop
three alternative perspectives on \emph{APL}'s contents, which they refer to
as the \textbf{generalized}, \textbf{creator}, and \textbf{user} perspectives \cite{Dawes2018}.
These perspectives amount to different techniques for \ignorelink[REORDERING KNOWLEDGE]{REORDERING
KNOWLEDGE}.  We will elaborate at the next level.

\subsection{System: Systemic approaches and solutions (social system)}
\label{sec:org7ef7d37}
The \textbf{system} layer is typically understood in terms of the \textbf{social
phenomena} that cause the problems at the litany layer to emerge (along
with their familiar solutions).  In the original setting in which
patterns developed, this layer would have included causes such as more
people living in cities, combined with the possibility of developing a
more community-driven approach to design using contemporary
technologies.  In short, at this level, we examine where the familiar problems come from.

Dawes and Ostwald’s \cite{Dawes2018} central finding is that many patterns in which
Alexander had medium or low confidence in fact occupy a relatively
central position in \emph{APL}'s graph:

\begin{quote}
\ldots{} the patterns which are most likely to be encountered by designers –
are most easily accessed, or provide greatest access to other patterns
– might be those which Alexander acknowledged were incapable of
providing fundamental solutions to the problems they addressed.
\end{quote}

This means that novice users could be expected to encounter problems
in application of \emph{APL}'s patterns: “despite its often authoritative and
dogmatic tone, Alexander’s text was framed as a work in progress,
rather than a definitive design guide” (p. 22).  Dawes and Ostwald
suggest that their analysis could point to “prime opportunities to
continue the development of \emph{A Pattern Language}'' (p. 21).

\begin{echo}
Here, a range of media issues begin to crop
up.\endnote{At this point it is also useful to recall that there are a
range of ‘other’ pattern discourses which could be relevant to
understanding how the problems emerge: here, ‘other’ is intended in
the sense mentioned in our \ignorelink[REORDERING KNOWLEDGE]{REORDERING KNOWLEDGE} pattern, i.e., pointing to other communities who are not in
close touch with PLoP: these include PurPLSoc and the world of
practicing architects.}
\end{echo}
Broadly put: there have been some attempts at
creating systematic archives of patterns \cite{koppeRepos,inventadoRepos}, but these
efforts haven’t always garnered significant buy-in.
Importantly, the first-ever Wiki was developed in connection with a
platform for developing, sharing, and revising pattern languages
\cite{cunningham2013a}.\footnote{\url{http://wiki.c2.com/?PeopleProjectsAndPatterns}}\textsuperscript{,}\footnote{\url{http://c2.com/ppr/}}
However, there was a distinction between the discussions and the finished patterns.  In the 2013 retrospective,
Ward Cunningham and coauthor Michael Mehaffy write:
\begin{quote}
The original wiki technology functioned in a direct open-source mode,
which allowed individuals to contribute small pieces to incrementally
improve the whole. (\emph{ibid.})
\end{quote}
This is true if by “open source” we understand what you see when you click Edit—but
the statement could be misleading relative to contemporary usage, which is often linked with
the Open Source Initiative’s definition, which centers on the premise that
“Open source doesn’t just mean access to the source code.”\footnote{\url{https://opensource.org/osd}}
On the \texttt{c2} wiki, licensing was restrictive. Discussions were to take place in “letters and replies” rather than revision or annotation of the published patterns; rights associated with the finished patterns were closely guarded.\footnote{\url{http://c2.com/ppr/titles.html}}\textsuperscript{,}\footnote{\url{http://c2.com/ppr/about/copyright.html}}

Although Wiki technology could in principle have been a site for
ongoing \ignorelink[DECONSTRUCTION]{DECONSTRUCTION} of patterns, this hadn’t happened on \texttt{c2}.
This is itself interesting and worth deconstructing a bit.  Notably,
there were only \emph{four} published “letters and replies”.\footnote{\url{http://c2.com/ppr/letters/index.html}} Unfortunately, we could not find
a public archive of the design patterns mailing list where further
discussions took place.  This suggests certain factors of contingency
in the development of the discourse.  Over the years, some of these concerns have been addressed—e.g.,
through the introduction of Federated Wikis and new licensing
models—and other issues and concerns came to the fore.

Dawes and Ostwald’s \cite{Dawes2018} remarks on multiple perspectives on
pattern languages resonate Jenifer Tidwell’s charges against the Gang
of Four:

\begin{quote}
\ldots{} the reality of a software artifact that the developer sees is not
the only one that's important.  What about the user's reality?  Why
has that been ignored in all the software patterns work that's been
done?  Isn't the user's experience the ultimate reason for designing a
building or a piece of software?  If that's not taken into account,
how can we say our building -- or our software -- is “good”? — “The Gang of Four Are Guilty”\footnote{\url{http://www.mit.edu/\~jtidwell/gof\_are\_guilty.html}}
\end{quote}

Notice that the \emph{user} of the designed artifact has entered the
story as a different figure from the user of the pattern language,
whom we met above.  Tidwell’s critique suggests at least a
couple \ignorelink[ALTERNATIVE PASTS AND FUTURES]{ALTERNATIVE PASTS AND FUTURES}: e.g., what if the end-user had been
placed at the center the whole time?  Alternatively, what if the
primary focus of patterns was to facilitate interaction between
different stakeholders?  The fact that Tidwell’s book
\cite{tidwell2010designing} and an essay by Jans Borchers \cite{borchers2008pattern}
which drew inspiration from her critique both have over
1000 citations on Google Scholar shows that Tidwell’s perspective has
been impactful.  To get a sense of how the pattern community may have
been informed by this critique—alongside other related trends and concerns—we can look at
how the Writers Workshops at PLoP have evolved over time.  In Table \ref{tabplop}, a
selection of titles of workshop sessions show how the focus of PLoP evolved from
primarily ‘programming’ oriented to a much broader contextual view
over time.  Indeed, by 2019, the focus is almost exclusively ‘contextual’.
The way the themes under discussion have evolved brings to mind the layers of CLA.

\begin{table}[htbp]
\caption{\label{tabplop}Evolution of PLoP Writers Workshop topics in selected years: CLA in the wild?}
\centering
\begin{tabular}{llll}
\textbf{1997} & \textbf{2011} & \textbf{2015} & \textbf{2019}\\
Architecture & Architecture & Pattern Writing & Group Architecture\\
Roles and Analysis & Design & Software Architecture \& Process & Culture\\
People and Process & Information & Cloud \& Security & Meta\\
Domain Specific Techniques & People & Innovation \& Analysis & Education\\
OO Techniques & Pedagogy & People \& Education & \\
Non-OO Techniques &  &  & \\
\end{tabular}
\end{table}

\rowcolors{2}{gray!25}{white}
\subsection{Worldview: ways of knowing and alternative discourse}
\label{sec:org03add48}

The next layer comprises \textbf{worldviews} (e.g., Alexander’s view that
“There is a central quality which is the root criterion of life and
spirit in a man, a town, a building, or a wilderness”).

The situation with licensing on \texttt{c2} is particularly interesting in
light of Alexander’s perspective that \emph{APL} was a “living language”.  In
principle, Wiki technology might have presented the opportunity to
realize this vision fully for the first time, in a virtual setting.
Wiki technology did become widely influential when it was combined
with a free content license on Wikipedia (originally GNU FDL, later
CC-By-SA).

Fast-forwarding to the present day, Christopher Alexander’s website
\texttt{patternlanguage.com} writes about \href{https://www.patternlanguage.com/membership/memberstour3-struggle.html}{The Struggle for People to be Free}—but it is not referencing freedom in the GNU sense.

In 1979 he was concerned: “Instead of being widely shared, the
pattern languages which determine how a town gets made becomes
specialized and private.”  In 2021, \emph{APL} itself is only legally
available for subscribers or for people who purchase a paper copy of
the book. (Or through a library!)  Of course, like many famous texts
it can also be obtained extra-legally for download as a PDF: but that
format does not afford downstream users the opportunity to collaborate
on the text’s further development.

Gabriel and Goldman talk about sharing and gift culture in their essay
\href{https://dreamsongs.com/MobSoftware.html}{Mob Software: The Erotic Life of Code}.\footnote{Notably, this essay was
presented as a keynote talk at the same programming conference where
Alexander had delivered his keynote, \cite{alexander1999a}, four years previously.}  This reference suggests
another reason why sharing knowledge in non-editable formats can be
problematic.  Their primary source on gift culture is Hyde
\cite{hyde2019gift}, who talks about what happens when exchange items
are taken out of the gift exchange culture and put in a museum:

\begin{quote}
A commodity is truly “used up” when it is sold because nothing about
the exchange assures its return.  The visiting sea captain may pay
handsomely for a Kula necklace, but because the sale removes it from
the circle, it wastes it, no matter the price.  Gifts that remain
gifts can support an affluence of satisfaction, even without numerical
abundance. (\emph{ibid.}, Chapter 1, p. 29)
\end{quote}

Gabriel and Goldman reference the open source community—but not the
free software community, so we will follow their usage here—as the
origin of Mob Software.

\begin{quote}
Because the open source proposition asked the crucial first question,
I include it in what I am calling “mob software,” but mob software
goes way beyond what open source is up to today. \cite{gabriel2000mob}
\end{quote}

That question is: “What if what once was scarce is now abundant?”  It
is well known that the PLoP conference series builds on this idea: it
includes shepherding and workshops \cite{gabriel2002a} as well as games,
informal gifts, and other measures that aim to create a sense of
psychological safety: all features that make PLoP a space where
‘failure’ is OK and even celebrated, as per Mob Software.  The essay
develops its own criticisms of open source, e.g., “the open-source
community is extremely conservative” and forking happens rarely.
(Five years later, with the creation of Git, a certain form of forking became more
typical.)  Resonating with Tidwell’s critique from above:

\begin{quote}
One difference between open source and mob software is that open
source topoi are technological while mob software topoi are people
centered.
\end{quote}

On a technical basis, Gabriel’s vision sounds a lot like today’s world
of \emph{microservices}.
While his vision hasn’t fully come to pass—for example there are still many
services with proprietary source code—nowadays many big companies
are also big proponents of open source.  Here we can notice that
Gabriel was employing a technique of imagining \ignorelink[ALTERNATIVE PASTS AND FUTURES]{ALTERNATIVE PASTS AND
FUTURES}, e.g., he imagined a future in which:

\begin{quote}
Mentoring circles and other forms of workshop are the mainstay of
software development education. There are hundreds of millions of
programmers.
\end{quote}

\afterpage{
\begin{longtable}{|p{\textwidth}|}
\caption{\label{tabone}Key observations from VanDrunen’s critique of Gabriel’s “Mob Software” essay}
\\
\hline
“Kauffman’s work is about a rediscovery of the sacred, and it amounts to a proposal of the laws of self-organization as a new deity”\\
“One thing we find in common with Lewis Thomas’s ants, Kauffman’s autocatalytic sets of proteins, and the agents inhabiting Sugarscape is that they all lack intelligence.”\\
“In other words, the rules given by Gabriel describe only the conforming aspect of group behavior. In reality, there is a tension between independent and conforming tendencies, and the flock patterns emerge from the interaction between the two.”\\
“His examples of ‘mob activity’\ldots{} the making of the Oxford English Dictionary, cathedral-building, and open source software discussed later—all had oversight, master-planning of some sort.”\\
“There are several distinct senses of ‘gift’ that lie behind these ideas, but common to each of them is the notation that a gift is a thing we do not get by our own efforts.” [Quoting Hyde \cite{hyde2019gift}.]\\
“Certainly proprietary code is shared property among those working in a corporate development team, but it is not common to the larger community of software developers and users.”\\
“A computer program is not like a poem or a dance in this way; if the programmer is not able to produce something parsable in the programming language or cannot fit the instructions together in a logical way, the program simply will not work.”\\
“Gabriel’s own experience may color his perception. He founded a software company that produced programs for Lisp development and which went bankrupt after 10 years.”\\
“Moreover, if Gabriel means to suggest that these programming languages or models could have made programming more accessible to the masses lacking technical skill, it is quite a dubious claim”\\
\hline
\end{longtable}}

We would like to dig somewhat deeper into the foundations of the
worldview that Gabriel puts forth in this essay. Usefully, an article
by VanDrunen “traces the source of Gabriel’s ideas by examining the
authorities he cites and how he uses them and evaluates their validity
on their own terms” \cite{vandrunenchristian}.  VanDrunen’s critique functions
as a (detailed) \ignorelink[DECONSTRUCTION]{DECONSTRUCTION} of the thinking behind Gabriel’s essay.
Some key excerpts appear in Table \ref{tabone}.
It is worth noting that this is by no means a complete critique.  As
an an example of one direction that we will not have time and space to
develop here, some applications of the concept of ‘gift culture’ have
been criticized as hegemonic in nature \cite{Mallard2019}: should we
expect pattern-theoretic, mob, or free/libre/open source software
culture to be immune from such concerns?  VanDrunen’s critique is
useful for our purposes not because they provide the last word, but
because this criticism points to the importance of considering the deeper layers
in developing a concept or approach.  There may also be conflicts at
these deeper layers.

It is also worth noting that mob software is but one of many diverse
visions of the future of programming \cite{postmodernProgramming}.  An
embrace of diverse perspectives seems to be a fundamental part of the
associated worldview.  After all, the
primary theoretical model of a computer is termed ``universal''.
Perhaps there is a bit of a paradox or double bind here, insofar as we embrace diverse
perspectives just as long as they are compatible with our core tenets.
For at least some pattern authors, these include “their love of programs
and programming” (\emph{ibid.}).  (On this last point, both VanDrunen and Gabriel
seem to agree.)

\subsection{Myths: metaphors and narratives (longer term change)}
\label{sec:orga5fc5bc}

Lastly, there are \textbf{myths or metaphors} (e.g., Alexander’s idea that the
architect’s work is done ‘for the glory of God’ (see Galle
\cite{GALLE2020345}) or his conception that ‘primitive’ dwellings
contain more life).  To emphasize, CLA does not dismiss myths in the
slightest: on the contrary, they are what drive the other layers.
Another term that is used to characterize this layer is \textbf{narratives}.
VanDrunen surfaced various concepts in Gabriel’s essay that would be
at home at this level, for example, the concept of duende that Gabriel
takes over from Garcia Lorca originally derives from \emph{dueño de casa},
the name of a certain kind of household spirit.  However, myth here
does not just refer to such entities, but to the most deeply held
beliefs and concepts that underlie worldviews.

One important narrative for the pattern discourse is in plain view
within the terminology of problems and solutions, which come from
mathematics and physics.  Alexander worked \emph{at the level of narrative}
to connect the patterns discourse to a scientific worldview, seeking a
sense of objectivity.  For example, in “The Atoms of Environmental
Structure”:

\begin{quote}
most designers \ldots{} say that the environment cannot be right or wrong
in any objective sense but that it can only be judged according to
criteria, or goals, or policies, or values, which have themselves been
arbitrarily chosen.  We believe this point of view is mistaken.
\end{quote}

Notice that, here, the discourse is positioned as different from the
mainstream of architecture.  The key differentiator is not the
language of problems and solutions, which would be familiar to anyone
with an engineering background; rather, but in a certain notion of
\emph{wholeness}.  Which notion of wholeness remains to be surfaced.
Quoting, again, from “The Atoms of Environmental Structure”, we get
some relevant background:

\begin{quote}
We believe that all values can be replaced by one basic value:
everything desirable in life can be described in terms of freedom of
people’s underlying tendencies. \ldots{} The environment should give free
rein to all tendencies; conflicts between people’s tendencies must be
eliminated.
\end{quote}

Historically, there are at least two major varieties of wholeness: one
that is based on progressive differentiation (e.g., unfolding from
substance, per Spinoza), and the other generated by interaction
between components (e.g., mutually reflecting monads, per Leibniz).
In support of these allusions, a quote of Alexander from \emph{The Nature of
Order} (\emph{TNO}): it “may be best if we redefine the concept of God in a
way that is more directly linked to the concept of ‘the whole.’”
\begin{echo}This sounds a lot like Spinoza!\endnote{We can obtain some useful \ignorelink[DISTANCE]{DISTANCE} by thinking about how different kinds
of wholeness are associated with different symbols. In terms of
metaphors, we have already encountered overt images like that of
Chartres cathedral.  If we allow ourselves to explore further afield,
other symbols of wholeness come to mind: these include the circle, the
cross—or potentially the cross inside a circle,
\begingroup\alch\symbol{"3B}\endgroup.
Related but more
elaborated symbols include the circle with a cross rising above it
(\varTerra) which is both the modern astronomical symbol for Earth and
also linked with the Carthusian order, and the Rod of Asclepius
(\Asclepius, for the deity associated with healing or making whole)—this last symbol sometimes being inter-confused with the Caduceus
(\Caduceus, the symbol of Hermes, the deity associated with mediation
of various forms, and also echoed in the planetary symbol for Mercury,
\begingroup\alch\symbol{"53}\endgroup).}
\end{echo}
Indeed, the pattern discourse appears to draw from \emph{both} major traditions of wholeness, while also
seeking to unite them.  We get the idea of unfolding in \emph{APL} and other
pattern languages that work in a top-down manner: however, we also get
the notion of patterns and principles that are generative of emergent
phenomena.

At this level, architecture and programming were seen, by Alexander
\cite{alexander1999a}, to unite: his questions for the computer
scientists to whom he was speaking point in the direction of
bio-hacking and nanotechnology (e.g., for molecular self-assembly)—at least at the allusive level.  The following quote suggests we have
embarked on a fruitful track by attempting to think at the deeper
layers of the pattern discourse:
\begin{quote}
Generative patterns work indirectly; they work on the underlying
structure of a problem (which may not be manifest in the problem)
rather than attacking the problem directly.\footnote{\url{https://wiki.c2.com/?GenerativePattern}}
\end{quote}

\begin{echo}
Clearly, another key metaphor—which also has a
generative aspect—is the metaphor of \emph{language}.\endnote{“... as
in the case of natural languages, the pattern language is
generative. It not only tells us the rules of arrangement, but shows
us how to construct arrangements - as many as we want - which satisfy
the rules.”—\url{https://wiki.c2.com/?GenerativePattern}, quoting from
\emph{The Timeless Way Of Building}, pp. 185-6.}
\end{echo}
The prominence of linguistic metaphors within DPL reminds us that
Alexander’s architectural oeuvre contains many traces of symbols associated
with Hermes: a deity associated with communication and mediation.
Through these reflections we gain some useful \ignorelink[DISTANCE]{DISTANCE}.
\begin{quote}
In the house, [Hermes’] place is at the door, protecting the
threshold\ldots{} He could be found around city gates, intersections, state
borders, and tombs (the gateways to the other world). \cite{benvenuto1993hermes}
\end{quote}
At the time when Hermes was actively embraced as a deity, in some
traditions he was paired with Hestia, the goddess of the hearth, whose
“domain was internal, the closed, the fixed, the inward” (\emph{ibid.}, here
and in quotes later in this paragraph).  The discourse around patterns
contains some aspects that move towards foundations (e.g., in the form
of fundamental principles, per \emph{TNO}).  Such foundations could be
associated with Hestia, whereas Hermes would be on the side of
generativity and mutation.  The dichotomy seems to repeat itself
within the \emph{TNO} principles themselves: recalling that ``focus'' is the Latin
term for the hearth, Strong Centers would align with Hestia, whereas
Hermes would align more with Deep Interlock and Ambiguity.  The
resolution of the two forces within pattern language—as a form—seems be a variation of these Nietzschean lines: “anything that is
becoming returns” (i.e., is discussable as pattern), and, “contingency
resolves itself into necessity” (i.e., the wholeness of generativity
ultimately recovers the wholeness of unfolding).

\begin{echo}
Our task in this section has been to situate Alexander’s thought
relative to the myths and symbols of wholeness; we’ve surfaced some of
the tensions and dynamics that exist at this level.  Relationships to some
other contemporary thinkers are discussed by Elsheshtawy \cite{Elsheshtawy2001},
in particular, a relationship to Piaget’s conception of operational wholeness is developed.
Alexander, for his part, professed ignorance of French Structuralist theory (quoted at \emph{ibid.})—in particular, of Barthes and Foucault, whom Inayatullah draws upon—and he tags Nietzsche
as a nihilist, while distinguishing his own work
as comparatively hopeful \cite{alexander1991perspectives}. For further
reflections on Nietzsche and wholeness, see \cite{bishop2020holistic}.
For more on Hestia and Hermes in an architectural
context, see \cite{springhestia}.
\end{echo}

\rowcolors{2}{white}{white}
\section{SUPPLEMENT: Case study: Planning “Season 1” for the Emacs Research Group}
\label{sec:org8a2d2cd}
\label{Case_study}

In the Emacs Research Group, we did a PAR at the end of every session,
in our (approximately weekly, two-hour) meetings from November 2020 to
the time of writing.\footnote{Data archived at
  \url{https://github.com/exp2exp/exp2exp.github.io}, with meeting
  notes and PARs indexed and viewable on the web at
  \url{https://exp2exp.github.io/erg}.}  This allowed us to track
progress, and to surface key issues and concerns.  For example,
bootstrapping needs related to scheduling and collaboration tools,
along with persistent questions about how best to go public, are
documented in our first PAR.  Every six weeks or so, we merged
selected bullet-points from the collected PARs into the CLA outline in
an intuitive way, depending on which section they seemed to fit best.
We elaborated those bullet points into a narrative form, which we
jointly revised to accommodate new data as time went on.  We also
began to develop TODO items that would make the next steps for this
seminar group both actionable and meaningful.  Additionally, we
connected these TODO items to design patterns collected in the
\emph{Peeragogy Handbook} \cite{peeragogy-handbook-long} (with ongoing
work appearing at \texttt{peeragogy.org}).  The TODO items typically
are not concrete objectives, but are, rather, descriptions of
anticipated patterns of behavior—here linked to \emph{bona fide}
design patterns.  Some new proto-patterns are named in all-caps.  To
refine these items into tasks that are concretely doable will require
further breakdown, refinement, and elaboration; furthermore, managing
the TODO list as a whole will require ongoing (re-)prioritization.
Typically, TODO items at the Litany level are more immediately
actionable, whereas those at the lower levels may take longer to
realise.

Paragraphs summarising the CLA are augmented with representative data from the
seminar sessions, and further broken down into next steps which are cross-referenced
with peeragogy design patterns, like \textsc{Roadmap}
\cite{peeragogy-handbook-long}.\footnote{See \url{http://peeragogy.org/top} for a
reworking of the \emph{Peeragogy Handbook} as a unified pattern language,
which extends the earlier presentation in \cite{patterns-of-peeragogy}.}
\begin{echo}
We include data points supporting the CLA from the PARs carried
out in our 1\textsuperscript{st}, 10\textsuperscript{th}, and 18\textsuperscript{th} sessions (marked with \textbf{(1)}, \textbf{(10)}, and \textbf{(18)} below).
\end{echo}
By the time of our fourth iteration of the larger
\(\mathrm{PAR}\rightarrow\mathrm{CLA}\) cycle, each section had
accumulated around 20-30 bullet points of supporting data at a similar level of
granularity.

\subsection{Understanding data, headlines, empirical world (short term change)}
\label{sec:org910aa7e}

We’ve made progress since we started with the raw themes of Research
on/in/with Emacs back in November 2020. We’ve met almost every week
since then, and interviewed some interesting and varied guests. We
have a clearer idea of what we want to talk about at the next
EmacsConf, and how we can be of service to researchers and Emacs
users. We have been using a workflow that helps us carefully review
progress, diagnose issues, and manage our energy.  We’re understanding
how research is done by doing it, and keeping careful track of the
process. If a session doesn’t go as well as hoped, we think about why
(especially the chair). The idea is that you know what the bomb is, so you can at least hope to defuse it later. We try to adapt gracefully to
circumstances as they evolve, without being reactive because we know
we will be back again next week and the week after, etc.

\paragraph{Representative supporting data}
\label{sec:org1218597}
\begin{itemize}
\item[\textbf{(1)}] \emph{Everyone shared a brief intro and ideas so we got to know each other}
\item[\textbf{(10)}] \emph{We’ve brainstormed a couple of options for getting out there: White-papers, Grants, Journal papers (very concrete)}
\item[\textbf{(18)}] \emph{Alex: My major intention was to meet you guys and learn something, wanting to reinforce existing knowledge of emacs and develop it further}
\end{itemize}

\paragraph{Next Steps}
\label{sec:orgba39c21}
\begin{center}
\begin{tabular}{ll}
Maintain plans for the next six months & \textsc{Roadmap}\\
Process the following points & \textsc{Scrapbook}\\
Keep doing PARs and CLAs & \textsc{Assessment}\\
Develop our intention-based workflow & \textsc{Forum}\\
Mesh with other ongoing activities elsewhere & \textsc{Cooperation}\\
\end{tabular}
\end{center}

\subsection{Systemic approaches and solutions (social system)}
\label{sec:orgbbc0e93}

If we tackle big enough projects, it will bring with it the need for
collaboration. (And we need to respect these other parties.) We like
to create tangible deliverables (e.g., journal articles). However, “If
we knew what the outcome was it wouldn’t be research” — therefore,
we’re focusing initially on research methods and design documents. So
far many of the stakeholders have to do with free software, open
communities, peer learning: all of this is part of a broader
initiative. All of them will need some degree of systematised
activities and documents. This is what we’re experimenting with; in
principle, we can provide meta-management workflows.  In the way we
work together we make sure to take account of emotions, not just a
time table. We’ve been experimenting with futures methods that help us
use the future intelligently: neither trying to do everything ‘live’,
nor overburdening the future with a bunch of plans that can’t be
realised.

\paragraph{Representative supporting data}
\label{sec:org8ead6aa}
\begin{itemize}
\item[\textbf{(1)}] \emph{Part of a greater sense of trying to do something with EmacsConf to federate the community}
\item[\textbf{(1)}] \emph{Joe: Leo did an amazing job facilitating the meeting}
\item[\textbf{(1)}] \emph{Public Policy conference: (How to get a grant?)}
\item[\textbf{(10)}] \emph{Potential interview with Leo \& Jethro Kuan (co-maintainers of org-roam)}
\item[\textbf{(18)}] \emph{Leo did a nice job of intervening}
\end{itemize}

\paragraph{Next Steps}
\label{sec:org97f52a9}
\begin{center}
\begin{tabular}{ll}
Identify potential stakeholders in Emacs Research & \textsc{Community}\\
Identify stakeholders in the kind of activities we can support & \textsc{A Specific Project}\\
Identify venues where we can reach these different stakeholders & \textsc{Wrapper}\\
Create some publication to plant a flag for our group & \textsc{Paper}\\
Keep exploring! & SERENDIPITY\\
\end{tabular}
\end{center}

\subsection{Worldview, ways of knowing and alternative discourse}
\label{sec:orge5529f3}
\label{erg_worldview}

We have looked at RStudio and Roam Research as models of (some of) the
kinds of things we think Emacs can eventually improve upon. Actually
getting there requires thinking about the specificity of what Emacs
can do. At least nominally, it is a system for editing. For example,
currently we can edit a wiki using Org Roam and Git. What about
editing distributed knowledge graphs? This would allow people to
reference ongoing research processes. Or we could go further and
contribute to the development of a new distributed read-write Web!
Alongside such software products could come various services, such as
a matchmaking service for academics, or a set of 24/7 virtual
conferences. To make such a thing really useful, we need to get
coherence out of various long-running, diverse, and heterogeneous
thought processes: and make sure we help people address real
problems. We can start small, working with the members of ERG and
their networks. We’re certainly not the only people who are struggling
with some unexpected commitments.  Successful adaptation requires the
articulation of an entire system. Thinking again about Emacs: its
current documentation is certainly useful, but it leaves many gaps,
some of which are filled in other ways (e.g., by mailing lists). As we
work we are paying attention to the growth not only of knowledge, but
also of capability. For this, we often rely on our feelings (getting
it, not getting it, accomplishing something or not, etc.).

\paragraph{Representative supporting data}
\label{sec:org9e91c2b}
\begin{itemize}
\item[\textbf{(1)}] \emph{Wonderful outcome from attending EmacsConf 2020!}
\item[\textbf{(10)}] \emph{Anthropology + Psychology is a special nightmare for reproducibility}
\item[\textbf{(10)}] \emph{Maybe the ERG could contribute further patterns?}
\item[\textbf{(18)}] \emph{But there’s a problem with Emacs, which is that there isn’t proper intro}
\end{itemize}

\paragraph{Next Steps}
\label{sec:org0a2ab9f}
\begin{center}
\begin{tabular}{ll}
Survey related work & \textsc{Context}\\
Spec out the Emacs based ‘answer’ to RStudio, Roam Research & \textsc{Community}\\
Continue to develop and refine our methods & \textsc{Assessment}\\
Product and business development plans for a multigraph interlinking service & \textsc{Website}\\
\end{tabular}
\end{center}

\subsection{Myths, metaphors and narratives: imagined (longer term change)}
\label{sec:org2f7f290}

In our concrete methods, we have aligned ourselves with the ‘long-term
perspective’. This includes both retrospective and prospective
thinking. For example, the things that were timely 7 years ago might
not be now; in many cases the relevance of a given innovation goes down over
time. That said, Emacs has an evolutionary character that has
allowed it to keep up with the times — apparently becoming
increasingly relevant and useful ever since Steele and Stallman
started to systematise Editor MACroS for the Text Editor and Corrector
(TECO) program. Not only has the technology evolved, but so has the
social setting in which this work is done. After nearly a year of
working together, we’re now prepared to argue that other people should
consider getting together to create their own communities similar to
ours! The concepts underlying the free software movement were based on
“communal sharing” of source code: now we’re working on developing and
sharing other methods as well. Just like free software has
legitimately expanded the range of what’s humanly possible, so too may
further efforts expand the frontier of possibility. As above, this takes
careful articulation, and a willingness to do truly original research
(without relying on fixed assumptions about what research is meant to
be, or where we will find value).

\paragraph{Representative supporting data}
\label{sec:orgf25be31}
\begin{itemize}
\item[\textbf{(1)}] \emph{We generally agreed that we want to make something that exposes intrinsic value of using these tools}
\item[\textbf{(10)}] [None recorded at this level from this PAR.]
\item[\textbf{(18)}] \emph{But there was no such guidance; you were in the middle of an alien playground. “But I just wanted to do my Clojure stuff.”}
\end{itemize}

\paragraph{Next Steps}
\label{sec:org16d024e}
\begin{center}
\begin{tabular}{ll}
Assess what we’re learning & \textsc{Assessment}\\
Think about how we can help improve gender balance in Free Software & DIVERSITY\\
\end{tabular}
\end{center}

\bibliographystyle{ACM-Reference-Format-Journals}

\clearpage
\end{document}